\documentclass{sig-alternate}
\pdfoutput=1

\usepackage[english]{babel}
\usepackage[utf8x]{inputenc}
\usepackage[T1]{fontenc}
\usepackage{lmodern}
\usepackage{flushend}

\usepackage{amscd}
\usepackage{latexsym}
\usepackage[usenames,dvipsnames]{color}
\usepackage{proof}

\usepackage{graphicx}
\graphicspath{{figures/}}

\usepackage{tikz}
\usepackage{array}

\usepackage{url}

\usepackage{amsmath,amsfonts,stackrel,stmaryrd,wasysym}
\usepackage{multirow}
\usepackage{graphicx}
\graphicspath{{figures/}}
\usepackage{color}

\newcommand{\Cc}{\mathcal{C}}
\newcommand{\Dc}{\mathcal{D}}

\newcommand{\Pc}{\mathcal{P}}

\newcommand{\Z}{\mathbb{Z}}

\newcommand{\op}[2]{\langle #1, #2 \rangle}

\newcommand{\forc}[3]{{\bf for} $#1$ := $#2$ {\bf to} $#3$}
\newcommand{\fore}{{\bf for each}}

\newcommand{\si}{{\bf if}}

\newcommand{\ia}{\ \ }
\newcommand{\ib}{\ \ \ \ }
\newcommand{\ic}{\ \ \ \ \ \ }

\newcommand{\vi}{\vec{i}}
\newcommand{\vj}{\vec{j}}

\newcommand{\thetat}[1]{\theta_{\hbox{\tiny #1}}}

\definecolor{ashgrey}{rgb}{0.7, 0.75, 0.71}
\newcommand{\gris}[1]{{\color{ashgrey}{#1}}}

\newcommand{\inorder}{\hbox{in-order}}
\newcommand{\unicity}{\hbox{unicity}}
\newcommand{\fifo}{\hbox{fifo}}
\newcommand{\dcc}{{\sc Dcc}}

\title{
Improving Communication Patterns in Polyhedral Process Networks}
\numberofauthors{1}
\author{
\alignauthor
Christophe Alias\\
\affaddr{CNRS, ENS de Lyon, Inria, UCBL, Universit\'e de Lyon}\\
\texttt{christophe.alias[at]ens-lyon.fr}
}

\begin{document}

\pagenumbering{arabic}
\toappear{
	\hrule \vspace{5pt}
	HIP3ES 2018\\
	Sixth International Workshop on High Performance Energy Efficient Embedded Systems\\
	Jan 22th, 2018, Manchester, UK\\
	In conjunction with HiPEAC 2018.\\[10pt]
	\url{https://www.hipeac.net/events/activities/7528/hip3es/}\\
}

\maketitle

\begin{abstract}
Embedded system performances are bounded by power consumption. The
trend is to offload greedy computations on hardware accelerators as
GPU, Xeon Phi or FPGA. FPGA chips combine both flexibility of
programmable chips and energy-efficiency of specialized hardware and
appear as a natural solution. 
Hardware compilers from high-level languages (High-level synthesis,
HLS) are required to exploit all the capabilities of FPGA while
satisfying tight time-to-market constraints.
%
Compiler optimizations for parallelism and data locality restructure
deeply the execution order of the processes, hence the read/write
patterns in communication channels. This breaks most FIFO channels,
which have to be implemented with addressable buffers. Expensive
hardware is required to enforce synchronizations, which often results
in dramatic performance loss.
In this paper, we present an algorithm to partition the communications
so that most FIFO channels can be recovered after a loop tiling, a key
optimization for parallelism and data locality. Experimental results
show a drastic improvement of FIFO detection for regular kernels at
the cost of a few additional storage. As a bonus, the
storage can even be reduced in some cases.
\end{abstract}

\section{Introduction}
\label{section:introduction}

Since the end of Dennard scaling, the performance of embedded systems
is bounded by power consumption. The trend is to trade genericity
(processors) for energy efficiency (hardware accelerators) by
offloading critical tasks to specialized hardware. FPGA chips combine
both flexibility of programmable chips and energy-efficiency of
specialized hardware and appear as a natural solution. High-level
synthesis (HLS) techniques are required to exploit all the
capabilities of FPGA, while satisfying tight time-to-market
constraints. 
Parallelization techniques from high-performance compilers are
progressively migrating to HLS, particularly the models and algorithms
from the polyhedral model \cite{pluto}, a powerful framework to
design compiler optimizations. Additional constraints must be
fulfilled before plugging a compiler optimization into an HLS
tool. Unlike software, the hardware size is bounded by the available
silicium surface. The bigger is a parallel unit, the less it can be
duplicated, thereby limiting the overall performance. Particularly,
tricky program optimizations are likely to spoil the performances if
the circuit is not post-optimized carefully \cite{alias_dags}. An
important consequence is that the the roofline model is not longer
valid in HLS \cite{dhollander13rooflinefpga}. Indeed, peak performance
is no longer a constant: it decreases with the operational
intensity. The bigger is the operational intensity, the bigger is the
buffer size and the less is the space remaining for the computation
itself. Consequently, it is important to produce at source-level a
precise model of the circuit which allows to predict accurately the
resource consumption.
Process networks are a natural and convenient intermediate
representation for HLS
\cite{dpn,rijpkema2000deriving,turjan2007phd,ppn2010chapter}. A
sequential program is translated to a process network by partitioning
computations into processes and flow dependences into channels. Then,
the processes and buffers are factorized and mapped to hardware.

In this paper, we focus on the translation of buffers to hardware. We
propose an algorithm to restructure the buffers so they can be mapped
to inexpensive FIFOs.  Most often, a direct translation of a regular
kernel -- without optimization -- produces to a process network with
FIFO buffers \cite{turjan2007classifying}.  Unfortunately, data
transfers optimization \cite{alias_date13} and generally loop tiling
reorganizes deeply the computations, hence the read/write order in
channels (communication patterns). Consequently, most channels may no
longer be implemented by a FIFO. Additional circuitry is required to
enforce synchronizations
\cite{dpn,zissulescu2002solving,turjan2002realizations,van2012enabling}
which result in larger circuits and causes performance penalties. In
this paper, we make the following contributions:
\begin{itemize}
\item We propose an algorithm to reorganize the communications between
  processes so that more channels can be implemented as FIFO after a
  loop tiling. As far as we know, this is the first algorithm to
  recover FIFO communication patterns after a compiler optimization.
\item Experimental results show that we can recover most of the FIFO
  disabled by communication optimization, and more generally any loop
  tiling, at almost no extra storage cost.
\end{itemize}

The remainder of this paper is structured as follows. Section
\ref{section:dpn} introduces polyhedral process network and discusses how
communication patterns are impacted by loop tiling, Section
\ref{section:algo} describes our algorithm to reorganize channels,
Section \ref{section:results} presents experimental results,
Finally, Section \ref{section:conclusion} concludes this paper and
draws future research directions.

\section{Preliminaries}
\label{section:dpn}

This section defines the notions used in the remainder of this paper.
Section \ref{section:poly} and \ref{section:schedule} introduces the
basics of compiler optimization in the polyhedral model and defines
loop tiling. Section \ref{section:ppn} defines polyhedral process
networks (PPN), shows how loop tiling disables FIFO communication
patterns and outlines a solution.


\subsection{Polyhedral Model at a Glance}
\label{section:poly}

\begin{figure*}[ht]
\begin{tabular}{lll}
  \begin{minipage}{0.45\textwidth}
    \begin{tabular}{l@{}l}
      &    \forc{i}{0}{N+1}\\
      $\bullet$&    \ia load($a[0,i]$);\\
      &    \forc{t}{1}{T}\\
      &    \ia \forc{i}{1}{N}\\
      $\gris{\bullet}$&    \ib $a[t,i] := a[t-1,i-1] + a[t-1,i] + a[t-1,i+1]$;\\
      &    \forc{i}{1}{N}\\
      $\circ$&    \ia store($a[T,i]$);\\
    \end{tabular}
  \end{minipage}
  &
  \begin{minipage}{0.2\textwidth}
    \includegraphics[scale=0.5]{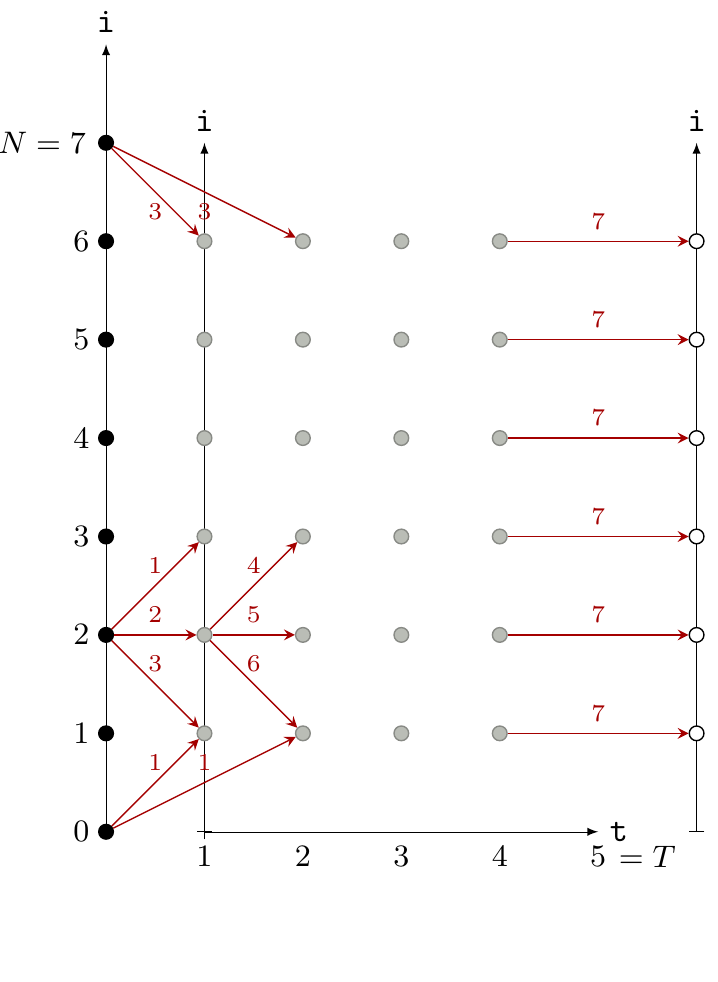}
  \end{minipage}
  &
  \begin{minipage}{0.2\textwidth}
    \includegraphics[scale=0.5]{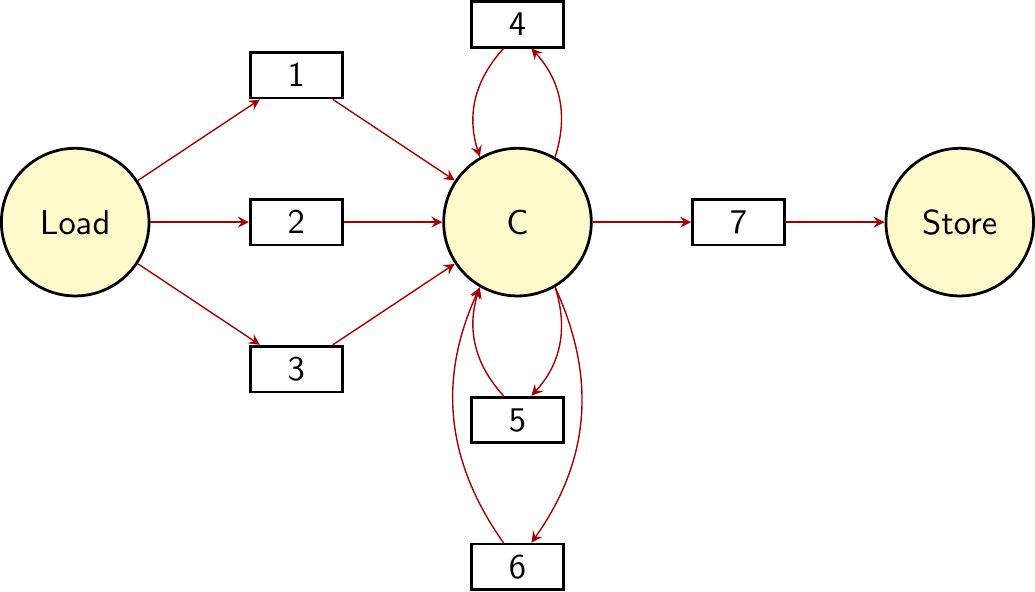}
  \end{minipage}
\\
(a) Jacobi 1D kernel  & (b) Flow dependences & (c) Polyhedral process network\\
\end{tabular}
\caption{\label{figure:jacobi1} Motivating example: Jacobi-1D kernel}
\end{figure*}

Translating a program to a process network requires to split the
computation into processes and flow dependences into channels.  The
{\em polyhedral model} focuses on kernels whose computation and flow
dependences can be predicted, represented and explored at
compile-time. The control must be predictable:
{\tt for} loops, {\tt if} with conditions on loop counters. Data
structures are bounded to arrays, pointers are not allowed. Also, loop
bounds, conditions and array accesses must be affine functions of
surrounding loop counters and structure parameters (typically the
array size). This way, the computation may be represented with
Presburger sets (typically approximated with convex polyhedra, hence
the name). This makes possible to reason geometrically about the
computation and to produce precise compiler analysis thanks to integer
linear programming: flow dependence analysis \cite{ada}, scheduling
\cite{pluto} or code generation \cite{cloog,quillere2000generation} to
quote a few.  Most compute-intensive kernels from linear algebra and
image processing fit in this category. In some cases, kernels with
dynamic control can even fit in the polyhedral model after a proper
abstraction \cite{alias_sas10}.
Figure \ref{figure:jacobi1}.(a) depicts a polyhedral kernel and (b)
depicts the geometric representation of the computation for each
assignment ($\bullet$ for assignment {\em load}, $\gris{\bullet}$ for
assignment {\em compute} and $\circ$ for assignment {\em store}). The
vector $\vi = (i_1, \ldots , i_n)$ of loop counters surrounding an
assignment $S$ is called an {\em iteration} of $S$. The execution of
$S$ at iteration $\vi$ is denoted by $\op{S}{\vi}$. The set $\Dc_S$ of
iterations of $S$ is called {\em iteration domain} of $S$. The
original execution of the iterations of $S$ follows the lexicographic
order $\ll$ over $\Dc_S$. For instance, on the statement $C$: $(t,i)
\ll (t',i')$ iff $t<t'$ or ($t=t'$ and $i<i'$). The lexicographic
order over $\Z^d$ is naturally partitioned by depth: $\ll = \ll^1
\uplus \ldots \uplus \ll^d$ where $(u_1 \ldots u_d) \ll^k (v_1,
\ldots, v_d)$ iff $\left( \wedge_{i=1}^{k-1} u_i = v_i\right) \wedge
u_k<v_k$.

\paragraph{Dataflow Analysis}
On Figure \ref{figure:jacobi1}.(b), red arrows depict several flow
dependences (read after write) between executions instances. We are
interested in flow dependences relating the production of a value to
its consumption, not only a write followed by a read to the same
location. These flow dependences are called {\em direct}
dependences. Direct dependences represent the communication of values
between two computations and drive communications and
synchronizations in the final process network. They are crucial to
build the process network. Direct dependences can be computed exactly
in the polyhedral model \cite{ada}. 
%
The result is a relation $\rightarrow$ relating each producer
$\op{P}{\vi}$ to one or more consumers $\op{C}{\vj}$. Technically,
$\rightarrow$ is a {\em Presburger relation} between vectors $(P,\vi)$
and vectors $(C,\vj)$ where assignments $P$ and $C$ are encoded as
integers. For example, dependence $5$ is summed up with the Presburger
relation: $\{ (\gris{\bullet}, t-1,i) \rightarrow (\gris{\bullet},t,i),\; 0<t\leq T \wedge 0 \leq i
\leq N \}$. Presburger relations are computable and efficient
libraries allow to manipulate them \cite{isl,omega}. In the remainder,
direct dependence will be referred as flow dependence or dependence to
simplify the presentation.

\subsection{Scheduling and Loop Tiling}
\label{section:schedule}
Compiler optimizations change the execution order to fulfill multiple
goals such as increasing the parallelism degree or minimizing the
communications. The new execution order is specified by a {\em
  schedule}. A schedule $\theta_S$ maps each execution $\op{S}{\vi}$
to a timestamp $\theta_S(\vi) = (t_1, \ldots, t_d) \in \Z^d$, the
timestamps being ordered by the lexicographic order $\ll$. In a way, a
schedule dispatches each execution instance $\op{S}{\vi}$ into a new
loop nest, $\theta_S(\vi) = (t_1, \ldots, t_d)$ being the new
iteration vector of $\op{S}{\vi}$. A schedule $\theta$ induces a new
execution order $\prec_\theta$ such that $\op{S}{\vi} \prec_\theta
\op{T}{\vj}$ iff $\theta_S(\vi) \ll \theta_T(\vj)$. Also, $\op{S}{\vi}
\preceq_\theta \op{T}{\vj}$ means that either $\op{S}{\vi}
\prec_\theta \op{T}{\vj}$ or $\theta_S(\vi) = \theta_T(\vj)$. When a
schedule is injective, it is said to be {\em sequential}: no execution
is scheduled at the same time. Hence everything is executed in
sequence. In the polyhedral model, schedules are affine
functions. They can be derived automatically from flow dependences
$\cite{pluto}$. On Figure \ref{figure:jacobi1}, the original execution
order is specified by the schedule $\thetat{load}(i) = (0,i)$,
$\thetat{compute}(t,i) = (1,t,i)$ and $\thetat{store}(i) = (2,i)$. The
lexicographic order ensures the execution of all the {\em load}
instances (0), then all the {\em compute} instances (1) and finally
all the {\em store} instances (2). Then, for each statement, the loops
are executed in the specified order.

{\em Loop tiling} is a transformation which partitions the computation
in tiles, each tile being executed atomically. Communication
minimization \cite{alias_date13} typically relies on loop tiling to
tune the ratio computation/communication of the program beyond the
ratio peak performance/communication bandwidth of the target
architecture. Figure \ref{figure:tiling}.(a) depicts the iteration
domain of {\em compute} and the new execution order after tiling loops
$t$ and $i$. For presentation reasons, we depict a domain bigger than
in Figure \ref{figure:jacobi1}.(b) (with bigger $N$ and $M$) and we
depict only a part of the domain. In the polyhedral model, a loop
tiling is specified by hyperplanes with linearly independent normal
vectors $\vec\tau_1, \ldots, \vec\vec\tau_d$ where $d$ is the number
of nested loops (here $\vec\tau_1 = (0,1)$ for the vertical
hyperplanes and $\vec\tau_2 = (1,1)$ for the diagonal
hyperplanes). Roughly, hyperplans along each normal vector
$\vec\tau_i$ are placed at regular intervals $b_i$ (here $b_1 = b_2 =
2$) to cut the iteration domain in tiles. Then, each tile is
identified by an iteration vector $(\phi_1, \ldots, \phi_d)$, $\phi_k$
being the slice number of an iteration $\vi$ along normal vector
$\vec\tau_k$: $\phi_k = \vec\tau_k \cdot \vi \div b_k$. The result is
a Presburger iteration domain, here $\hat\Dc = \{
(\phi_1,\phi_2,t,i),\; 2\phi_1 \leq t < 2(\phi_1+1) \wedge 2\phi_2
\leq t+i < 2(\phi_2+1) \}$: the polyhedral model is closed under loop
tiling. In particular, the tiled domain can be scheduled. For
instance, $\hat\theta_S(\phi_1,\phi_2,t,i) = (\phi_1,\phi_2,t,i)$
specifies the execution order depicted in Figure
\ref{figure:tiling}.(a)): tile with point (4,4) is executed, then tile
with point (4,8), then tile with point (4,12), and so on. For each
tile, the iterations are executed for each $t$, then for each $i$.

\subsection{Polyhedral Process Networks}
\label{section:ppn}
Given the iteration domains and the flow dependence relation,
$\rightarrow$, we derive a {\em polyhedral process network} by
partitioning iterations domains into processes and flow dependence into
channels. More formally, a polyhedral process network is a couple
$(\Pc,\Cc)$ such that:
\begin{itemize}
\item
Each process $P \in \Pc$ is specified by an iteration domain $\Dc_P$
and a sequential schedule $\theta_P$ inducing an execution order
$\prec_P$ over $\Dc_P$. Each iteration $\vi \in \Dc_P$ realizes the
execution instance $\mu_P(\vi)$ in the program. The processes
partition the execution instances in the program: $\{ \mu_P(\Dc_P)$
for each process $P\}$ is a partition of the program computation.
\item
Each channel $c \in \Cc$ is specified by a producer process $P_c \in
\Pc$, a consumer process $C_c \in \Pc$ and a dataflow relation
$\rightarrow_c$ relating each production of a value by $P_c$ to its
consumption by $C_c$: if $\vi \rightarrow_c \vj$, then execution $\vi$
of $P_c$ produces a value read by execution $\vj$ of $C_c$.
$\rightarrow_c$ is a subset of the flow dependences from $P_c$ to
$C_c$ and the collection of $\rightarrow_c$ for each channel $c$
between two given processes $P$ and $C$, $\{
\rightarrow_c,\; (P,C) = (P_c,C_c) \}$, is a partition of flow
dependences from $P$ to $C$.
\end{itemize}
The goal of this paper is to find out a partition of flow dependences
for each producer/consumer couple $(P,C)$, such that most channels
from $P$ to $C$ can be realized by a FIFO.

Figure \ref{figure:jacobi1}.(c) depicts the PPN obtained with the
canonical partition of computation: each execution $\op{S}{\vi}$ is
mapped to process $P_S$ and executed at process iteration $\vi$:
$\mu_{P_S}(\vi) = \op{S}{\vi}$. For presentation reason the {\em
  compute} process is depicted as $C$.  Dependences depicted as $k$ on
the dependence graph in (b) are solved by channel $k$.  To read the
input values in parallel, we use a different channel per couple
producer/read reference, hence this partitioning. We assume that, {\em
  locally}, each process executes instructions in the same order than
in the original program: $\thetat{load}(i) = i$,
$\thetat{compute}(t,i) = (t,i)$ and $\thetat{store}(i) = i$. Remark
that the leading constant (0 for {\em load}, 1 for {\em compute}, 2
for {\em store}) has disappeared: the timestamps only define an order
local to their process: $\prec_{load}$, $\prec_{compute}$ and
$\prec_{store}$. The global execution order is driven by the dataflow
semantics: the next process operation is executed as soon as its
operands are available. The next step is to detect communication
patterns to figure out how to implement channels.

\paragraph{Communication Patterns}
A channel $c \in \Cc$ might be implemented by a FIFO iff the consumer
$C_c$ read the values from $c$ in the same order than the producer
$P_c$ write them to $c$ ({\em in-order}) and each value is read
exactly once ({\em unicity})
\cite{turjan2007phd,turjan2007classifying}. The {\em in-order}
constraint can be written:
\[
\begin{array}{l}
\inorder(\rightarrow_c, \prec_P, \prec_C) := \\
\qquad \forall x \rightarrow_c x', \forall y \rightarrow_c y': x' \prec_C y'
\Rightarrow x \preceq_P y
\end{array}
\]

The unicity constraints can be written:
\[
\begin{array}{l}
\unicity(\rightarrow_c) := \\
\qquad \forall x \rightarrow_c x', \forall y \rightarrow_c y': x' \neq y'
\Rightarrow x \neq y
\end{array}
\]
Notice that unicity depends only on the dataflow relation
$\rightarrow_c$, it is independent from the execution order of the
producer process $\prec_P$ and the consumer process $\prec_C$.
Furthermore, $\neg \inorder(\rightarrow_c, \prec_P, \prec_C)$ and
$\neg \unicity(\rightarrow_c)$ amount to check the emptiness of a
convex polyhedron, which can be done by most LP solvers.


Finally, a channel may be implemented by a FIFO iff it verifies
both in-order and unicity constraints:
\[
\begin{array}{l}
\fifo(\rightarrow_c, \prec_P, \prec_C) :=\\
\qquad \inorder(\rightarrow_c, \prec_P, \prec_C) \wedge \unicity(\rightarrow_c)
\end{array}
\]
When the consumer reads the data in the same order than they are
produced but a datum may be read several times:
$\inorder(\rightarrow_c, \prec_P, \prec_C) \wedge \neg
\unicity(\rightarrow_c)$, the communication pattern is said to be {\em
  in-order with multiplicity}: the channel may be implemented with a
FIFO and a register keeping the last read value for multiple
reads. However, additional circuitry is required to trigger the write
of a new datum in the register \cite{turjan2007phd}: this
implementation is more expensive than a single FIFO. Finally, when we
have neither $\inorder$ nor $\unicity$: $\neg \inorder(\rightarrow_c,
\prec_P, \prec_C) \wedge \neg \unicity(\rightarrow_c)$, the
communication pattern is said to be {\em out-of-order without
  multiplicity}: significant hardware resources are required to
enforce flow- and anti- dependences between producer and consumer and
additional latencies may limit the overall throughput of the circuit
\cite{dpn,zissulescu2002solving,turjan2002realizations,van2012enabling}.

\begin{figure*}[ht]
\begin{center}
\begin{tabbing}
\=
      \begin{minipage}{0.2\textwidth}
        \includegraphics[scale=0.5]{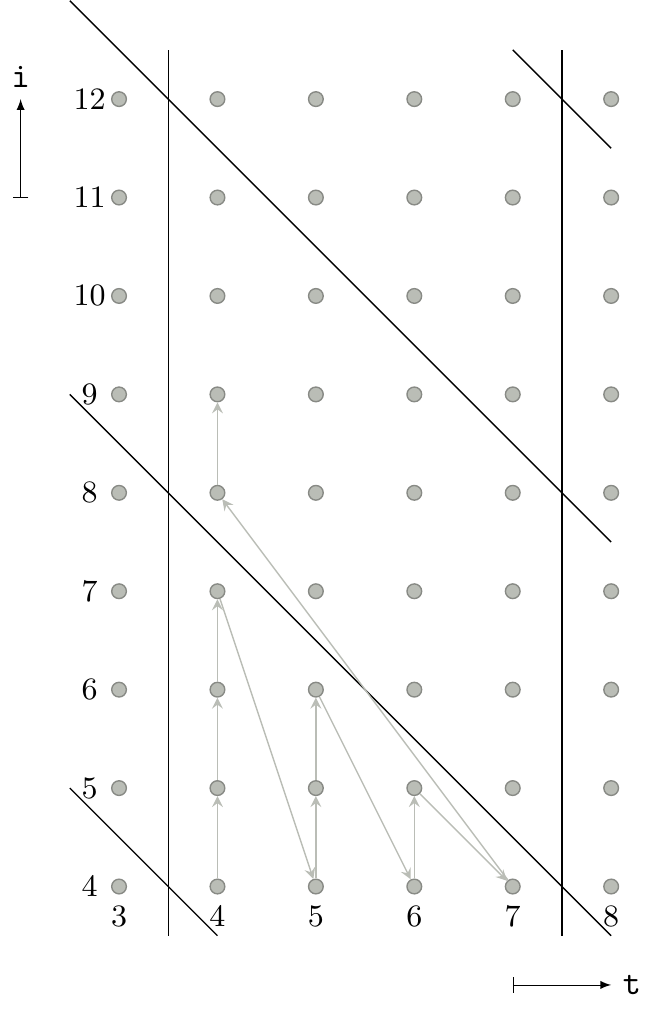}
      \end{minipage}
\hspace{1cm}
\=
      \begin{minipage}{0.2\textwidth}
        \includegraphics[scale=0.5]{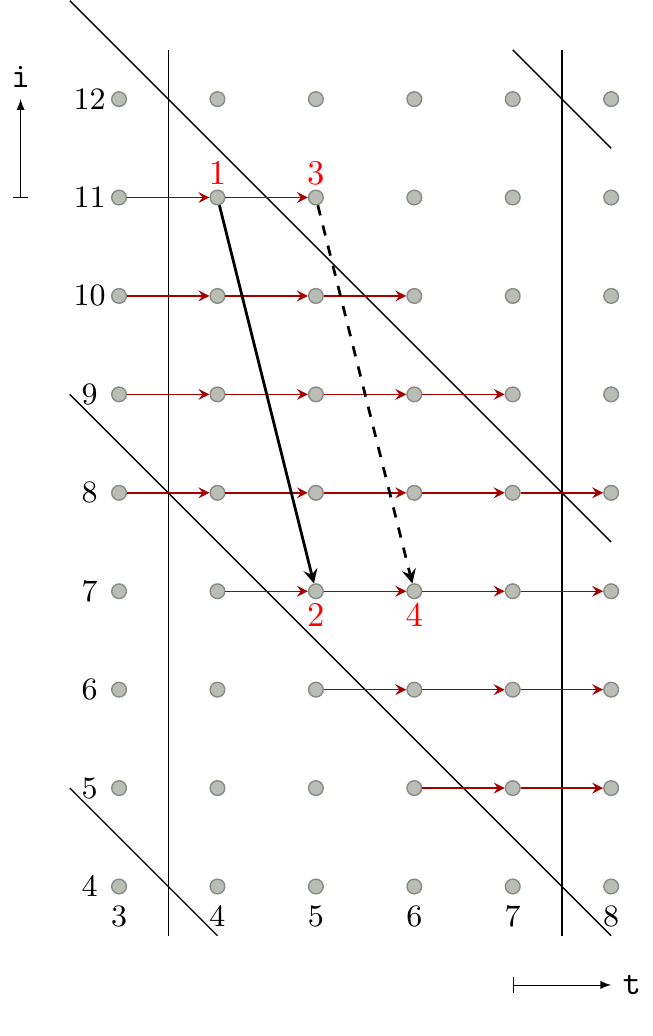}
      \end{minipage}
\hspace{1cm}
\=
      \begin{minipage}{0.2\textwidth}
        \includegraphics[scale=0.5]{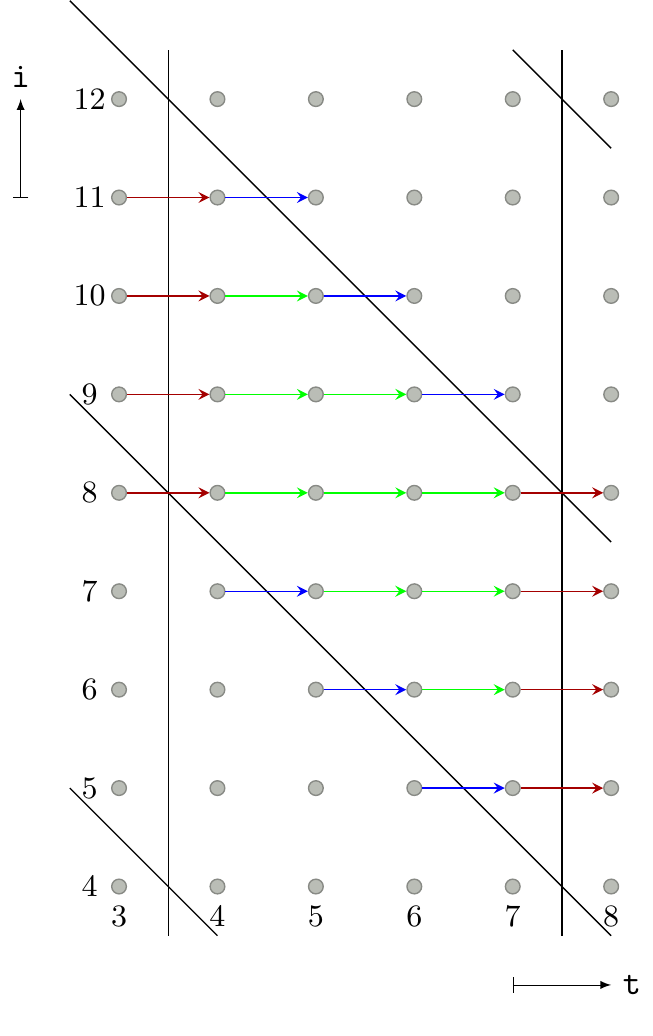}
      \end{minipage}
\hspace{1cm}
\=
      \begin{minipage}{0.2\textwidth}
        \includegraphics[scale=0.5]{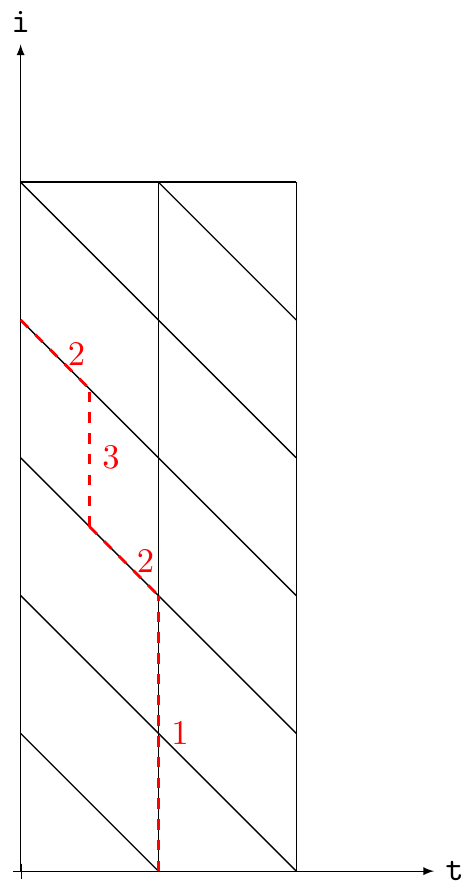}
      \end{minipage}
\\
\> (a) Loop tiling \> (b) Communication pattern \> (c) Our solution \> (d) Storage requirements
    \end{tabbing}
  \end{center}
  \caption{\label{figure:tiling}Impact of loop tiling on
    communication pattern}
\end{figure*}

Consider Figure \ref{figure:jacobi1}.(c), channel $5$, implementing
dependence $5$ (depicted on (b)) from $\op{\gris{\bullet}}{t-1,i}$
(write $a[t,i]$) to $\op{\gris{\bullet}}{t,i}$ (read $a[t-1,i]$).
With the schedule defined above, the data are produced
($\op{\gris{\bullet}}{t-1,i}$) and read ($\op{\gris{\bullet}}{t-1,i}$)
in the same order, and only once: the channel may be implemented as a
FIFO. Now, assume that process {\em compute} follows the tiled
execution order depicted in Figure \ref{figure:tiling}.(a).  The
execution order now executes tile with point (4,4), then tile with
point (4,8), then tile with point (4,12), and so on.  In each tile,
the iterations are executed for each $t$, then for 
\begin{figure}[ht]
{\small
\begin{tabular}{ll}
{\tiny 1} & {\sc split}($\rightarrow_c$,$\theta_P$,$\theta_C$)\\
{\tiny 2} & \ia \forc{k}{1}{n}\\
{\tiny 3} & \ib {\sc add}($\rightarrow_c \cap \{ (x,y),\; \theta_P(x) \ll^k \theta_C(y) \}$); \\
{\tiny 4} & \ia {\sc add}($\rightarrow_c \cap \{ (x,y),\; \theta_P(x) \approx^n \theta_C(y) \}$);\\
\\
{\tiny 5} & {\sc fifoize}($(\Pc,\Cc)$)\\
{\tiny 6} & \ia {\fore} channel $c$\\
{\tiny 7} & \ib $\{ \rightarrow_c^1, \ldots, \rightarrow_c^{n+1}\}$ := {\sc split}($\rightarrow_c$,$\theta_{P_c}$,$\theta_{C_c}$);\\
{\tiny 8} & \ib {\si} fifo($\rightarrow_c^k,\prec_{\theta_{P_c}},\prec_{\theta_{C_c}}$) $\forall k$\\
{\tiny 9} & \ic {\sc remove}($\rightarrow_c$); \\
{\tiny 10} & \ic {\sc insert}($\rightarrow_c^k$) $\forall k$;
\end{tabular}}
\caption{\label{figure:algo}Our algorithm for partitioning channels}
\end{figure}

\noindent each $i$. Consider
iterations depicted in red as $1,2,3,4$ in Figure
\ref{figure:tiling}.(b). With the new execution order, we execute
successively 1,2,4,3, whereas an in-order pattern would have required
1,2,3,4. Consequently, channel $5$ is no longer a FIFO. The same hold
for channel 4 and 6. Now, the point is to partition dependence $5$ and
others so FIFO communication pattern hold.

Consider Figure \ref{figure:tiling}.(c). Dependence $5$ is
partitioned in 3 parts: red dependences crossing tiling hyperplane
$\phi_1$ (direction $t$), blue dependences crossing tiling hyperplane
$t+i$ (direction $t+i$) and green dependences inside a tile. Since
the execution order in a tile is the same than the original execution
order (actually a subset of the original execution order), green
dependences will clearly verify the FIFO communication pattern. As
concerns blue and red dependences, source and target are executed in
the same order because the execution order is the same for each tile
and dependence $5$ happens to be short enough. In practice, this
partitioning is effective to reveal FIFO channels. In the next
section, we propose an algorithm to find such a partitioning.

\section{Our Algorithm}
\label{section:algo}

Figure \ref{figure:algo} depicts our algorithm for partitioning
channels given a polyhedral process network $(\Pc,\Cc)$ (line 5).  For
each channel $c$ from a producer $P=P_c$ to a consumer $C =C_c$, the
channel is partitioned by depth along the lines described in the
previous section (line 7).  $\Dc_P$ and $\Dc_C$ are assumed to be
tiled with the same number of hyperplanes. $P$ and $C$ are assumed to
share a schedule with the shape: $\theta(\phi_1,\ldots, \phi_n,\vi) =
(\phi_1,\ldots, \phi_n,\vi)$. This case arise frequently with tiling
schemes for I/O optimization \cite{dpn}. If not, the next channel
$\rightarrow_c$ is considered (line 6). The split is realized by
procedure {\sc split} (lines 1--4). A new partition is build starting
from the empty set. For each depth (hyperplane) of the tiling, the
dependences crossing that hyperplane are filtered and added to the
partition (line 3): this gives dependences $\rightarrow_c^1, \ldots,
\rightarrow_c^n$. Finally, dependences lying in a tile (source and
target in the same tile) are added to the partition (line 4): this
gives $\rightarrow_c^{n+1}$. $\theta_P(x) \approx^n \theta_C(y)$ means
that the $n$ first dimensions of $\theta_P(x)$ and $\theta_C(y)$
(tiling coordinates ($\phi_1, \ldots, \phi_n$)) are the same: $x$ and
$y$ belong to the same tile.
Consider the PPN depicted in Figure \ref{figure:jacobi1}.(c) with the
tiling and schedule discussed above: process {\em compute} is tiled
as depicted in Figure \ref{figure:tiling}.(c) with the schedule
$\thetat{compute}(\phi_1,\phi_2,t,i) = (\phi_1,\phi_2,t,i)$.  Since
processes {\em load} and {\em store} are not tiled, the only channels
processed by our algorithm are 4,5 and 6. {\sc split} is applied on
the associated dataflow relations $\rightarrow_4$, $\rightarrow_5$ and
$\rightarrow_6$. Each dataflow relation is split in three parts as
depicted in Figure \ref{figure:tiling}.(c). For $\rightarrow_5$:
$\rightarrow_5^1$ crosses hyperplane $t$ (red), $\rightarrow_5^2$
crosses hyperplane $t+i$ (blue) and $\rightarrow_5^3$ stays in a tile
(green).

This algorithm works pretty well for short uniform dependences
$\rightarrow_c$: if $\fifo(c)$ before tiling, then, after tiling, the
algorithm can split $c$ in such a way that we get FIFOs. However, when
dependences are longer, {\em e.g.} $(t,i) \rightarrow (t,i+2)$, the
target operations $(t,i+2)$ reproduce the tile execution pattern,
which prevents to find a FIFO. The same happens when the tile
hyperplanes are ``too skewed'', e.g. $\tau_1=(1,1)$, $\tau_2=(2,1)$,
dependence $(t-1,i-1) \rightarrow (t,i)$.
Figure \ref{figure:tiling}.(d) depicts the volume of data to be stored
on the FIFO produced for each depth. In particular, dotted line with
$k$ indicates iterations producing data to be kept in the FIFO at
depth $k$. FIFO at depth 1 (dotted line with 1) must store $N$ data at
the same time. Similarly, FIFO at depth 2 stores at most $b_1$ data
and FIFO at depth 3 stores at most $b_2$ data. Hence, on this example,
each transformed channel requires $b_1+b_2$ additional storage. In
general the additional storage requirements are one order of magnitude
smaller than the original FIFO size and stays reasonable in practice,
as shown in the next section.
\section{Experimental Evaluation}
\label{section:results}

This section presents the experimental results obtained on the
benchmarks of the polyhedral community. We demonstrate the
capabilities of our algorithm at recovering FIFO communication
patterns after loop tiling and we show how much additional storage is
required.

\paragraph{Experimental Setup}
We have run our algorithm on the kernels of PolyBench/C v3.2
\cite{polybench}. Tables \ref{table:results:fifo} and
\ref{table:results:size} depicts the results obtained for each kernel.
Each kernel is tiled to reduce I/O while exposing parallelism
\cite{dpn} and translated to a PPN using our research compiler, {\dcc}
(DPN C Compiler). {\dcc} actually produces a DPN (Data-aware Process
Network), a PPN optimized for a specific tiled pattern. DPN features
additional control processes and synchronization for I/O and
parallelism which have nothing with our optimization. So, we actually
only consider the PPN part of our DPN. We have applied our algorithm
to each channel to expose FIFO patterns. For each kernel, we compare
the PPN obtained after tiling to the PPN processed by our algorithm.

\paragraph{Results}
Table \ref{table:results:fifo} depicts the capabilities of our
algorithm to find out FIFO patterns. For each kernel, we provide the
channels characteristics on the original tiled PPN (Before
Partitioning) and after applying our algorithm (After
Partitioning). We give the total number of channels (\#channel), the
FIFO found among these channels (\#fifo), the number of channels which
were successfully turned to FIFO thanks to our algorithm
(\#fifo-split), the ratios \#fifo/\#channel (\%fifo) and
\#fifo-split/\#channel (\%fifo-split), the cumulated size of the FIFO
found (fifo-size) and the cumulated size of the channels found,
including FIFO (total-size).
On every kernel, our algorithm succeeds to expose more FIFO patterns
(\%fifo vs \%fifo-split). On a significant number of kernels (11 among
15), we even succeed to turn all the compute channels to FIFO. On the
remaining kernels, we succeed to recover all the FIFO communication
patterns disabled by the tiling. Even though our method is not
complete, as discussed in section \ref{section:algo}, it happens that
all the kernels fulfill the conditions expected by our algorithm
(short dependence, tiling hyperplanes not too skewed).

Table \ref{table:results:size} 
depicts the additional storage required
after splitting channels. For each kernel, we compare the cumulative
size of channels split and successfully turn to a FIFO
(size-fifo-fail) to the cumulative size of the FIFOs generated by the
splitting (size-fifo-split). The size unit is a datum {\em e.g.} 4
bytes if a datum is a 32 bits float. We also quantify the additional
storage required by split channels compared to the original channel
($\Delta$ := [size-fifo-split - size-fifo-fail] / size-fifo-fail). 
It turns out that the FIFO generated by splitting use mostly the same
data volume than the original channels. Additional storage resources
are due to our sizing heuristic \cite{bee}, which rounds channel size
to a power of 2.
Surprisingly, splitting can sometimes help the sizing heuristic to
find out a smaller size (kernel {\tt gemm}), and then reducing the
storage requirements. Indeed, splitting decomposes a channel into
channels of a smaller dimension, for which our sizing heuristic is
more precise. In a way, our algorithm allows to find out a nice
piecewise allocation function whose footprint is smaller than a single
piece allocation. We plan to exploit this nice side effect in the
future.

\begin{table}[!h]
\begin{center}
\begin{tabular}{|l|rrr|}
\hline
kernel & size-fifo-fail & size-fifo-split & $\Delta$\\
\hline
trmm & 256 & 257 & 0\%\\
gemm & 512 & 288 & -44\%\\
syrk & 8192 & 8193 & 0\%\\
symm & 800 & 801 & 0\%\\
gemver & 32 & 33 & 3\%\\
gesummv & 0 & 0 & \\
syr2k & 8192 & 8193 & 0\%\\
lu & 528 & 531 & 1\%\\
cholesky & 273 & 275 & 1\%\\
atax & 1 & 1 & 0\%\\
doitgen & 4096 & 4097 & 0\%\\
jacobi-2d & 8320 & 8832 & 6\%\\
seidel-2d & 49952 & 52065 & 4\%\\
jacobi-1d & 1152 & 1174 & 2\%\\
heat-3d & 148608 & 158992 & 7\%\\
\hline
\end{tabular}

\end{center}
\caption{Impact on storage requirements}\label{table:results:size}
\end{table}

\break
\section{Conclusion}
\label{section:conclusion}

In this paper, we have proposed an algorithm to reorganize the
channels of a polyhedral process network to reveal more FIFO
communication patterns. Specifically, our algorithm operates
producer/consumer processes whose iteration domain has been partitioned
by a loop tiling. Experimental results shows that our algorithm allows
to recover the FIFO disabled by loop tiling with almost the
same storage requirement.
Our algorithm is sensible to the dependence size and the chosen loop
tiling. In the future, we plan to design a reorganization algorithm
provably complete, in the meaning that a FIFO channel will be
recovered whatever the dependence size and the tiling used. We also
observe that splitting channels can reduce the storage requirements in
some cases. We plan to investigate how such cases can be revealed
automatically.

\begin{table*}[ht]
\begin{center}
{\small
\begin{tabular}{|l|rrrrrrr|rrrr|}
\hline
\multirow{2}{*}{Kernel} & \multicolumn{7}{l|}{Before Partitioning} & \multicolumn{4}{l|}{After Partitioning} \\
 & \#channel & \#fifo & \#fifo-split & \%fifo & \%fifo-split & fifo-size & total-size & \#channel & \#fifo & fifo-size & total-size\\
\hline
trmm & 2 & 1 & 2 & 50\% & 100\% & 256 & 512 & 3 & 3 & 513 & 513\\
gemm & 2 & 1 & 2 & 50\% & 100\% & 16 & 528 & 3 & 3 & 304 & 304\\
syrk & 2 & 1 & 2 & 50\% & 100\% & 1 & 8193 & 3 & 3 & 8194 & 8194\\
symm & 6 & 3 & 6 & 50\% & 100\% & 18 & 818 & 7 & 7 & 819 & 819\\
gemver & 6 & 3 & 5 & 50\% & 83\% & 4113 & 4161 & 7 & 6 & 4146 & 4162\\
gesummv & 6 & 6 & 6 & 100\% & 100\% & 96 & 96 & 6 & 6 & 96 & 96\\
syr2k & 2 & 1 & 2 & 50\% & 100\% & 1 & 8193 & 3 & 3 & 8194 & 8194\\
lu & 8 & 0 & 3 & 0\% & 37\% & 0 & 1088 & 11 & 6 & 531 & 1091\\
cholesky & 9 & 3 & 6 & 33\% & 66\% & 513 & 1074 & 11 & 8 & 788 & 1076\\
atax & 5 & 3 & 4 & 60\% & 80\% & 48 & 65 & 5 & 4 & 49 & 65\\
doitgen & 3 & 2 & 3 & 66\% & 100\% & 8192 & 12288 & 4 & 4 & 12289 & 12289\\
jacobi-2d & 10 & 0 & 10 & 0\% & 100\% & 0 & 8320 & 18 & 18 & 8832 & 8832\\
seidel-2d & 9 & 0 & 9 & 0\% & 100\% & 0 & 49952 & 16 & 16 & 52065 & 52065\\
jacobi-1d & 6 & 1 & 6 & 16\% & 100\% & 1 & 1153 & 10 & 10 & 1175 & 1175\\
heat-3d & 20 & 0 & 20 & 0\% & 100\% & 0 & 148608 & 38 & 38 & 158992 & 158992\\
\hline
\end{tabular}
}
\end{center}
\caption{\label{table:results:fifo} Detailed results}
\end{table*}

\bibliographystyle{abbrv}
\bibliography{main}

\begin{thebibliography}{10}

\bibitem{bee}
C.~Alias, F.~Baray, and A.~Darte.
\newblock {Bee+Cl@k}: An implementation of lattice-based array contraction in
  the source-to-source translator {Rose}.
\newblock In {\em ACM Conf. on Languages, Compilers, and Tools for Embedded
  Systems (LCTES'07)}, 2007.

\bibitem{alias_sas10}
C.~Alias, A.~Darte, P.~Feautrier, and L.~Gonnord.
\newblock Multi-dimensional rankings, program termination, and complexity
  bounds of flowchart programs.
\newblock In {\em International Static Analysis Symposium (SAS'10)}, 2010.

\bibitem{alias_date13}
C.~Alias, A.~Darte, and A.~Plesco.
\newblock Optimizing remote accesses for offloaded kernels: Application to
  high-level synthesis for {FPGA}.
\newblock In {\em ACM SIGDA Intl. Conference on Design, Automation and Test in
  Europe (DATE'13)}, Grenoble, France, 2013.

\bibitem{dpn}
C.~Alias and A.~Plesco.
\newblock {Data-aware Process Networks}.
\newblock Research Report RR-8735, {Inria - Research Centre Grenoble --
  Rh{\^o}ne-Alpes}, June 2015.

\bibitem{alias_dags}
C.~Alias and A.~Plesco.
\newblock {Optimizing Affine Control with Semantic Factorizations}.
\newblock {\em {ACM Transactions on Architecture and Code Optimization (TACO)
  }}, 14(4):27, Dec. 2017.

\bibitem{cloog}
C.~Bastoul.
\newblock Efficient code generation for automatic parallelization and
  optimization.
\newblock In {\em 2nd International Symposium on Parallel and Distributed
  Computing {(ISPDC} 2003), 13-14 October 2003, Ljubljana, Slovenia}, pages
  23--30, 2003.

\bibitem{pluto}
U.~Bondhugula, A.~Hartono, J.~Ramanujam, and P.~Sadayappan.
\newblock A practical automatic polyhedral parallelizer and locality optimizer.
\newblock In {\em Proceedings of the {ACM} {SIGPLAN} 2008 Conference on
  Programming Language Design and Implementation, Tucson, AZ, USA, June 7-13,
  2008}, pages 101--113, 2008.

\bibitem{dhollander13rooflinefpga}
B.~da~Silva, A.~Braeken, E.~H. D'Hollander, and A.~Touhafi.
\newblock Performance modeling for {FPGAs}: extending the roofline model with
  high-level synthesis tools.
\newblock {\em International Journal of Reconfigurable Computing}, 2013:7,
  2013.

\bibitem{ada}
P.~Feautrier.
\newblock Dataflow analysis of array and scalar references.
\newblock {\em International Journal of Parallel Programming}, 20(1):23--53,
  1991.

\bibitem{omega}
W.~Kelly, V.~Maslov, W.~Pugh, E.~Rosser, T.~Shpeisman, and D.~Wonnacott.
\newblock The omega calculator and library, version 1.1. 0.
\newblock {\em College Park, MD}, 20742:18, 1996.

\bibitem{polybench}
L.-N. Pouchet.
\newblock Polybench: The polyhedral benchmark suite.
\newblock {\em URL: http://www. cs. ucla. edu/\~{}
  pouchet/software/polybench/[cited July,]}, 2012.

\bibitem{quillere2000generation}
F.~Quiller{\'e}, S.~Rajopadhye, and D.~Wilde.
\newblock Generation of efficient nested loops from polyhedra.
\newblock {\em International journal of parallel programming}, 28(5):469--498,
  2000.

\bibitem{rijpkema2000deriving}
E.~Rijpkema, E.~F. Deprettere, and B.~Kienhuis.
\newblock Deriving process networks from nested loop algorithms.
\newblock {\em Parallel Processing Letters}, 10(02n03):165--176, 2000.

\bibitem{turjan2007phd}
A.~Turjan.
\newblock {\em Compiling nested loop programs to process networks}.
\newblock PhD thesis, Leiden Institute of Advanced Computer Science (LIACS),
  and Leiden Embedded Research Center, Faculty of Science, Leiden University,
  2007.

\bibitem{turjan2002realizations}
A.~Turjan, B.~Kienhuis, and E.~Deprettere.
\newblock Realizations of the extended linearization model.
\newblock {\em Domain-specific processors: systems, architectures, modeling,
  and simulation}, pages 171--191, 2002.

\bibitem{turjan2007classifying}
A.~Turjan, B.~Kienhuis, and E.~Deprettere.
\newblock Classifying interprocess communication in process network
  representation of nested-loop programs.
\newblock {\em ACM Transactions on Embedded Computing Systems (TECS)}, 6(2):13,
  2007.

\bibitem{van2012enabling}
S.~van Haastregt and B.~Kienhuis.
\newblock Enabling automatic pipeline utilization improvement in polyhedral
  process network implementations.
\newblock In {\em Application-Specific Systems, Architectures and Processors
  (ASAP), 2012 IEEE 23rd International Conference on}, pages 173--176. IEEE,
  2012.

\bibitem{isl}
S.~Verdoolaege.
\newblock {ISL}: An integer set library for the polyhedral model.
\newblock In {\em ICMS}, volume 6327, pages 299--302. Springer, 2010.

\bibitem{ppn2010chapter}
S.~Verdoolaege.
\newblock {\em {P}olyhedral {P}rocess {N}etworks}, pages 931--965.
\newblock Handbook of {S}ignal {P}rocessing {S}ystems. 2010.

\bibitem{zissulescu2002solving}
C.~Zissulescu, A.~Turjan, B.~Kienhuis, and E.~Deprettere.
\newblock Solving out of order communication using {CAM} memory: an
  implementation.
\newblock In {\em 13th Annual Workshop on Circuits, Systems and Signal
  Processing (ProRISC 2002)}, 2002.

\end{thebibliography}

\end{document}